\documentclass[aps,twocolumn,showpacs]{revtex4}
\usepackage{graphicx}   
\usepackage{latexsym}   

\begin{document}

\title{
Optimal conditions for exploring high-density baryonic matter}
\author{J.\ Randrup$^a$ and J.\ Cleymans$^b$}

\affiliation{$^a$Nuclear Science Division, 
Lawrence Berkeley National Laboratory, Berkeley, California 94720, USA}
\affiliation{$^b$Department of Physics, University of Cape Town, South Africa}

\date{\today}

\begin{abstract}
Using simple parametrizations of the thermodynamic freeze-out parameters
extracted from the data over a wide beam-energy range,
we reexpress the hadronic freeze-out line in terms of the underlying
dynamical quantities, the net baryon density $\rho_B$
and the energy density $\varepsilon$,
which are subject to local conservation laws.
This analysis makes it apparent that $\rho_B$ exhibits
a maximum as the collision energy is decreased.
This maximum freeze-out density has $\mu=400-500\,{\rm MeV}$,
which is above the critical value, and it is reached
for a fixed-target bombarding energy of $20-30\,{\rm GeV/A}$.
\end{abstract}


\maketitle


In recent years, it has become abundantly clear that the collision energy 
plays an important role in determining the properties of the final state
in relativistic heavy ion collisions.
While the extracted freeze-out temperature generally increases
monotonically with the collision energy,
the corresponding net baryon density exhibits a more complicated behavior:
It initially increases steadily as the collision energy is raised but
ultimately, at sufficiently high energies, 
it decreases as a result of the nuclear transparency.
Therefore, there is a certain beam energy for which
the resulting baryon density at freeze-out acquires a maximum value.
In this paper, we discuss this optimal beam energy
on the basis of the most up-to-date results 
on the properties of the hadronic freezeout.

For this purpose, we employ statistical considerations
which have been quite successful in accounting for the observed yields
of most hadronic species
(see Refs.\ \cite{Heinz,Becattini,MMRS,Rafelski,MRS}, for example).
In fact, the evidence has become truly overwhelming in recent years
that chemical equilibrium is established in relativistic heavy ion collisions 
throughout the entire range of collision energies .
The resulting picture shows that the freeze-out temperature $T$ 
increases steadily with the collision energy
as the corresponding baryon chemical potential $\mu_B$
decreases steadily towards zero.

However, it is not always the most convenient to work with
the thermodynamic variables $T$ and $\mu_B$.
This is particularly true within a dynamical context,
since they are not subject to any conservation laws,
in contrast to the more basic dynamical variables,
namely the energy density $\varepsilon$ 
and the net baryon density $\rho_B$.
Furthermore, when a first-order phase transition is present,
$T$ and $\mu_B$ become non-monotonic functions of $\varepsilon$ and $\rho_B$
throughout the associated phase-coexistence region of the phase diagram
(as does the pressure $p$).
This feature in turn causes the inverse functions to be multivalued,
a clearly inconvenient situation.
It is therefore of interest to reexpress the thermodynamic 
freeze-out information in terms of the dynamical quantities.
The resulting representation also serves to more clearly bring out the fact
that the freeze-out density exhibits a maximum value,
a feature that may be important in the planning of experiments
that seek to explore compressed baryonic matter.


In the grand-canonical ensemble of hadron species $i$,
the partition function factorizes into separate contributions for each specie,
${\cal Z}=\Pi_i {\cal Z}_i$, with
\begin{equation}
\ln{\cal Z}_i(T,V,\{\mu\}) = \pm{VTg_im_i^2\over2\pi^2}\sum_{n=1}^\infty
{(\pm\lambda_i)^{n}\over n^2} K_2({nm_i\over T})\ ,
\end{equation}
where plus is for bosons and minus is for fermions.
The hadron mass is $m_i$ and $g_i$=$2J_i$+$1$ is the spin degeneracy.
The fugacity associated with the hadron specie $i$ is
\begin{equation}
\lambda_i(T,\{\mu\})={\rm e}^{(\mu_BB_i+\mu_QQ_i+\mu_SS_i)/T}\ ,
\end{equation}
where $B_i$, $Q_i$, and $S_i$ 
denote its baryon number, electric charge,and strangeness, respectively.
The ensemble is characterized by the temperature $T$ and
the three chemical potentials $\{\mu\}=\{\mu_B,\mu_Q,\mu_S\}$.
Finally, $V$ denotes the enclosing volume of the system
(which plays no role in the analysis since only yield ratios are considered).

The corresponding spatial number density of a given specie is
then readily obtained,
\begin{eqnarray}\label{n1}
n_i(T,\{\mu\})\! &=&\! \pm{g_i T^3\over2\pi^2}\sum_{n=1}^\infty
{(\pm\lambda_i)^n\over n^3}  ({nm_i\over T})^2 K_2({nm_i\over T})~\ \\ 
\label{n2}
&=& {g_i\lambda_i\over2\pi^2}m_i^2 T K_2({m_i\over T})\ \pm\ \dots\ .
\end{eqnarray}
In the first line, the factors are arranged
such that the terms in the sum are regular in the small-mass limit, $m_i\to0$,
where $K_2$ diverges,
and the second line exhibits the leading term 
representing the classical result.
Since the sign of the second term in (\ref{n1}) 
depends on the quantum-statistical nature of the specie,
the leading term overestimates fermion densities
while it underestimates boson densities.
For all the parameter values employed in the yield estimates,
the first term in the expansion, Eq.\ (\ref{n2}),
is an excellent approximation for all baryon species,
being accurate to within a few per mille.
For kaons it is off by a few per cent,
and even the pion densities are underestimated by at most $10\%$.
The first term (\ref{n2}) would thus be a quite reasonable approximation
for rough estimates of the yields.

Once we know the number density of each specie,
it is straightforward to calculate the corresponding
baryon, charge, and strangeness densities.
In particular, the net baryon density is
\begin{equation}\label{rho}
\rho_B(T,\{\mu\})\ =\ T{\partial\ln{\cal Z}\over\partial\mu_B}\ 
=\ \sum_i B_i n_i(T,\{\mu\})\ .
\end{equation}
Furthermore, the total energy density is $\varepsilon=\sum_i\varepsilon_i$,
where the contribution by a particular specie $i$ is
\begin{eqnarray}\label{eps1}
\varepsilon_i(T,\{\mu\})\! &=&\! -{\partial\ln{\cal Z}_i\over\partial\beta}\
=\ \pm{g_i T^4\over2\pi^2}\sum_{n=1}^\infty
{(\pm\lambda_i)^n\over n^4}  \\ \nonumber
&\times& \left[({nm_i\over T})^2 K_2({nm_i\over T})
+({nm_i\over T})^3 K_1({nm_i\over T})\right]\\ \label{eps2}
&\approx& {g_i\lambda_i\over2\pi^2} m_i^2T^2 \left[K_2({m_i\over T})
+{m_i\over T} K_1({m_i\over T})\right]\ ,
\end{eqnarray}
where the last line is the classical result.

As it turns out, it is possible to fit the observed yield ratios
reasonably well with the statistical model,
by adjusting the Lagrange multipliers at each particular collision energy
\cite{Heinz,Becattini,MMRS,Rafelski,MRS}.
In this analysis, $T$ and $\mu_B$ are treated as free parameters,
while the values of $\mu_S$ and $\mu_Q$ are fixed 
by requiring overall strangenss neutrality, $\langle S\rangle=0$
and that the ratio of the net charge to the net baryon number
be equal to that of the collision system,
{\em e.g.}\ $\langle Q\rangle=0.4\langle B\rangle$ for Au+Au.
As discussed in Ref.\ \cite{Cleymans05},
the effect of the small change in this ratio in going to Pb+Pb is negligible.
The extracted values of the freeze-out temperature can be approximately
represented as \cite{Cleymans05},
\begin{equation}\label{T}
T(\mu_B)\ \approx\ 166 - 139\mu_B^2-53\mu_B^4\ ,
\end{equation}
where $T$ and $\mu_B$ are expressed in MeV.

\begin{figure}[tbh]
\includegraphics[angle=0,width=3.0in]{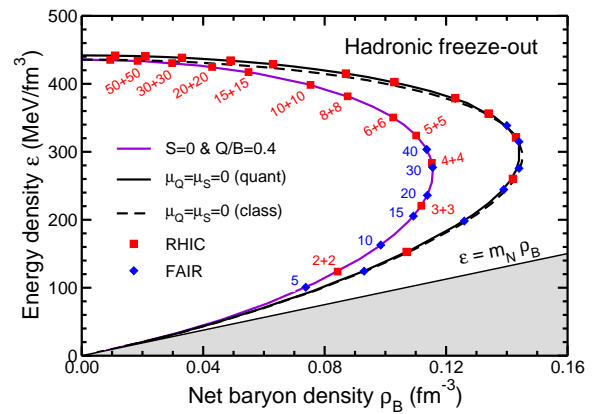}	
\caption{The hadronic freeze-out line in the $\rho_B-\varepsilon$ phase plane 
as obtained in the statistical model with the values of $\mu_B$ and $T$ that
have been extracted from the experimental data in Ref.\ \cite{Cleymans05}.
The curves on the right have been calculated for $\mu_Q=\mu_S=0$
using either quantal (solid) or classical (dashed) statistics, 
while the curve on the left employs values of $\mu_Q$ and $\mu_S$ 
that have been adjusted to ensure $\langle S\rangle=0$ 
and $\langle Q\rangle=0.4\langle B\rangle$ for each value of $\mu_B$.  
Also indicated are the beam energies (in GeV/$A$)
for which the particular freeze-out conditions are expected at
either RHIC (total energy in each beam),
starting at 100+100 and going down to 2+2,
or FAIR (kinetic energy of the beam for a stationary target),
starting at 5 and going up to 40.
as based on fits to existing data \cite{Cleymans05}.  
The triangular area corresponds to energy densities 
below the minimum required at the given net baryon density, 
$\varepsilon=m_N\rho_B$ (ignoring binding and compression),
and is thus inaccessible.
}\label{f:1}\end{figure}

Using the above relationship (\ref{T}),
we have calculated the corresponding freeze-out values of 
the net baryon density $\rho_B$ (Eq.\ (\ref{rho}))
and the energy density $\varepsilon$ (Eq.\ (\ref{eps1})).
The result is displayed in Fig.\ \ref{f:1}.
Before turning to the calculated results, 
we note that the lower-right section of the $\rho_B-\varepsilon$ plane
is not physically accessible, 
since a given net baryon density $\rho_B$ gives rise to 
a certain minimum energy density.
(The corresponding curve, $\varepsilon_{\rm min}(\mu_B)$,
is simply the pressure at zero temperature and is therefore occasionally,
and somewhat misleadingly, referred to as the equation of state.)
In the ideal-gas scenario, where binding and compression effects are absent,
this lower bound is given by $\varepsilon_{\rm min}=m_N\rho_B$,
which is indicated on the figure.

In order to illustrate the importance of adjusting $\mu_Q$ and $\mu_S$
to ensure average  conservation of charge and strangeness (see above),
we have also calculated those values of $\rho_B$ and $\varepsilon$
that would result if $\mu_Q$ and $\mu_S$ were taken to vanish.
As can be seen from Fig.\ \ref{f:1}, the failure to adjust $\mu_Q$ and $\mu_S$
increases the values of both densities by amounts 
that are epsecially significant in the region of maximal density.
These results were calculated both clasically (first terms only)
and quantally (all terms) in order to illustrate 
how unimportant this distinction is for the present analysis.

At the highest energies, 
freeze-out occurs for a negligible value of the chemical potential $\mu_B$,
(and hence the net baryon density $\rho_B$ is practically zero)
and the energy density $\varepsilon$ is nearly one half $\rm GeV/fm^3$.
As the collision energy is lowered, 
$\rho$ increases rapidly while $T$ initially remains fairly constant
but gradually begins to drop.
Then, in the range of $T=140-130\,{\rm MeV}$,
the freeze-out line $(\rho_B,\varepsilon)$ 
exhibits a backbend and approaches the origin.
The resulting maximum value of the net baryon density at freeze-out 
is about three quarters of the normal nuclear saturation density 
of $\rho_0\approx0.15\,{\rm fm}^{-3}$.

The value of the baryon chemical potential $\mu_B$ extracted from the data
decreases monotonically with the collision energy and can be parametrized
on a simple form \cite{Cleymans05},
\begin{equation}\label{mu}
\mu_B(\sqrt{s})\ \approx\ {1308 \over 1000+ 0.273\sqrt{s}}\ ,
\end{equation}
where $\mu_B$ as well as the NN CM energy $\sqrt{s}$ are expressed in MeV.
By use of this result, 
it is possible to attach beam energies along the freeze-out curves
shown in Fig.\ \ref{f:1}.
This has been done for both collider experiments,
such as those being carried out at RHIC, or for a fixed target,
as has been done at the SPS and the AGS and is being planned at FAIR.

We note that if this region of maximum freeze-out density
were to be explored with RHIC,
the appropriate beam kinetic energies would be about $2-4\,{\rm GeV}/A$,
corresponding to $s=(6-10\,{\rm GeV}/A)^2$,
which would be a rather challenging proposition.
By contrast, it appears that the region would be well within reach of FAIR
which is planned to bombard fixed targets with heavy-ion beams 
having kinetic energies of $5-40\,{\rm GeV/A}$.

Finally, according to the most refined lattice QCD studies 
\cite{Fodor:2001pe,Fodor:2004nz,Allton:2005gk,Gavai:2004sd,Lombardo},
it appears that the region of highest freeze-out density 
lies well beyond the critical region where the transformation between the
quark-gluon plasma and the hadronic resonance gas changes from being
merely a crossover to being a genuine first-order phase transition.
This fact is helpful to the prospect of probing this phase transition
by means of heavy-ion collisions.


Let us briefly summarize:
Analyses of experimentally obtained hadronic yield ratios
at a variety of collision energies have shown that the data
can be well reproduced within the conceptually simple statistical model
that describes an ideal hadron resonance gas in statistical equilibrium.
Furthermore, the extracted freeze-out values of the temperature and
the baryon chemical potential exhibit a smooth and monotonic
dependence on the collision energy and can be simply parametrized.
We have used these results to examine how the freeze-out appears
when represented in terms of the basic baryon and energy densities,
rather than chemical potential and temperature.
These quantities are more basic 
and they are of more direct relevance to the collision dynamics,
because they are subject to corresponding conservation laws.

We have found that the freeze-out value of the net baryon density
exhibits a maximum as the collision energy is being scanned,
thus suggesting that there may be an optimal collision energy (range)
for the exploration of compressed baryonic matter.
In a fixed-target configuration, this optimal beam kinetic energy is
$20-30\,{\rm GeV/A}$, which appears to be ideal for the planned FAIR at GSI,
and it may be accessible in the low-energy campain now being planned at RHIC.

~\\
This work was supported by the Office of Energy Research,
Office of High Energy and Nuclear Physics,
Nuclear Physics Division of the U.S. Department of Energy
under Contract No.\ DE-AC03-76SF00098.


                        \end{document}